\documentclass[11pt]{article}
 \usepackage{amsfonts}
\usepackage{amsmath} 
 \usepackage{amssymb}
 \def\bdt{\dot \beta}
 \def\adt{\dot \alpha}
 
 \newfont{\bbbold}{msbm10 scaled \magstep1}

 \def\cD{{\cal D}}
 
 \def\cF{{\cal F}}

 \def\cL{{\cal L}}
 \def\cM{{\cal M}}

 \newfont{\goth}{eufm10 scaled \magstep1}

 \def\a{\alpha}
 \def\b{\beta}
 \def\c{\gamma}

 \def\d{\delta}

 \def\e{\epsilon}

\def\L{\Lambda}
 \def\m{\mu}

 \def\s{\sigma}

 \def\t{\tau}
 \def\th{\theta}

 \def\del{\partial}

 \def\ua{\underline{\alpha}}

 \def\una{\underline a}
\def\unA{\underline A}
 \def\unb{\underline b}
\def\unB{\underline B}
 \def\unc{\underline c}
\def\unC{\underline C}
 \def\und{\underline d}

\def\unF{\underline F}
\def\unK{\underline K}
\def\unZ{\underline Z}
\def\unL{\underline L}
\def\unW{\underline W}

\def\unM{\underline M}

 \def\unH{\underline{H}}
\def\unG{\underline{G}}
 \def\unF{\underline{F}}

 \def\del{\partial}

 \def\be{\begin{equation}}
\def\ee{\end{equation}}
 \def\bea{\begin{eqnarray}}
\def\eea{\end{eqnarray}}
 \def\ba{\begin{array}}
\def\ea{\end{array}}

 \def\det{{\rm det\,}}

\def\lim{ \underset{\e \rightarrow 0}{\text{lim}}}
\def\intMo{\underset{\cM_o}{\int}}
 
\begin{document}

 \thispagestyle{empty}

 \hfill{KCL-MTH-01-48}

 \hfill{hep-th/0112048}

 \hfill{\today}

 \vspace{20pt}

 \begin{center}
 {\Large{\bf Relating Superembeddings and Non-linear Realisations}}
 \vspace{30pt}

 {J.M. Drummond}
\vskip 1cm {Department of Mathematics} \vskip 1cm {King's College,
London} \vspace{15pt}

 \vspace{60pt}

 {\bf Abstract}

 \end{center}
We discuss the relation between the superembedding method for deriving worldvolume actions
for D-branes and the method of Partially Broken Global Supersymmetry based upon linear and non-linear realisations of 
SUSY. We give the explicit relation for the cases of space filling
branes in 3 and 4 dimensions and show that the standard $\cF$-constraint 
of the superembedding method is the source of the required covariant non-linear constraints 
for the PBGS method.

{\vfill\leftline{}\vfill \vskip  10pt

 \baselineskip=15pt \pagebreak \setcounter{page}{1}


\section{Introduction}
A superembedding is the embedding of one superspace inside another. The theory of such embeddings provides an excellent
geometrical framework to describe the dynamics of superbranes.
For a review of superembeddings see \cite{d.s}. 
 In fact all BPS branes are described by a superembedding
satisfying a natural geometrical condition, namely that at all points on the brane
the odd tangent space of the brane is a subspace of the odd tangent
space of the superspacetime in which the brane lives. 

This `embedding condition' is often enough to describe the dynamics 
of the brane, that is it leads to a worldvolume multiplet for which an action can be written or sometimes it implies the 
equations of motion directly.
For some cases, including some D-branes, the embedding condition leads to an under-constrained multiplet. In these cases an 
additional constraint is required to enable one to construct the brane action or equations of motion. This constraint is
called the $\cF$-constraint.
In the D-brane cases, one introduces an independent worldvolume 2-form modified field strength $\cF$ satisfying $d\cF = - H$
where H is the pullback of a closed target space (Neveu-Schwarz) 3-form. One then constrains $\cF$ to have only purely
bosonic components. This constraint can be justified by considering branes ending on other branes \cite{cs,chs,chsw}.
With this additional constraint one then has either an off-shell multiplet, in which case a brane 
action can be written, or an on-shell multiplet, i.e. satisfying the equations of motion.   

In recent papers \cite{codzero, b.p.p.s.t} superembeddings with bosonic codimension zero were discussed in three, four and six 
dimensions. Although the number of bosonic dimensions is the same for the worldvolume and the target space these are genuine 
embeddings because the numbers of fermionic dimensions of these spaces are not the same. Specifically, the worldvolume has
exactly half the number of the target space since these branes preserve half the supersymmetry of the background supergravity.
The superembedding formalism was shown to be a powerful systematic method for the derivation of the 
dynamics of these space filling branes. In \cite{codzero} Green-Schwarz and superfield actions were constructed for the three
dimensional case. In \cite{codzero, b.p.p.s.t} the Green-Schwarz action in the four dimensional case 
was constructed and shown to be of the standard Born-Infeld type.

Since these branes preserve half the background supersymmetry, one can think of the superembedding of a space filling brane into 
flat space as the 
partial breaking of supersymmetry by one half (PBGS) \cite{pst} . The broken supersymmetry is said to be realised non-linearly.      
This idea has been applied to many cases \cite{bg,bg2,bik,bik2,ik,rt,dik} . 
In \cite{bg} the case of the space filling 3-brane was considered, and in \cite{ik} the membranes in three 
and four dimensions.
Firstly one introduces a multiplet which transforms linearly under the full
supersymmetry and then imposes constraints on it which relate its components to the Goldstone field of the broken 
supersymmetry. 
The superfield Lagrangian can be identified with one of the $N=1$ superfield components of the original multiplet. 
For the space filling branes in three and four dimensions
the multiplet which transforms linearly is the $N=2$ Maxwell multiplet. 
This multiplet satisfies a deformation of the standard Maxwell constraints.
In \cite{Iv}, an algorithm was given for deriving the constraints which relate the components of the $N=2$ Maxwell multiplet to the 
Goldstone field in these cases. 
The constraints were found using the relationship between linear and non-linear realisations of supersymmetry \cite{IvKa}, which was 
adapted to PBGS cases in \cite{dik,iklz}.

The relationship between non-linear realisations and superembeddings has been discussed before \cite{acghn,pst,pcw}. In \cite{pcw}
it was shown that the embedding condition of the superembedding formalism is equivalent to what is called the `inverse Higgs constraint'
of the non-linear realisation formalism. For the cases under consideration here the embedding condition can be imposed
without loss of generality and we focus on the relationship of the $\cF$-constraint to the non-linear realisations framework.
In this paper we explicitly show the equivalence between superembeddings and linear and non-linear realisations of PBGS 
for the cases of the space filling branes in three and four dimensions. We introduce a target space 2-form $\unF$ for
the $N=2$ Maxwell multiplet and we write the deformation of the $N=2$ Maxwell constraints as the modified Bianchi
identity $d\unF = - \unH$. Pulling this equation back to the worldvolume, we obtain the worldvolume Bianchi identity for the two form
$\cF$ introduced in the superembedding formalism.
We show that the non-linear constraints on the Maxwell fields imposed in the PBGS method are exactly the 
standard $\cF$-constraint. This equivalence allows us
to show that the superfield actions defined by the superembedding method are the same as those invariants constructed using
non-linear realisations of supersymmetry. 

We begin by reviewing very briefly the procedure of embedding space filling branes into three and four dimensional flat 
superspaces and describing the method for the construction of brane actions. We then go on to show the equivalence with
the linear and non-linear realisations method in the three dimensional case and then the four dimensional case. In section 3
we show the equivalence of the constraints of the two methods. In section 4 we show the equivalence of the actions defined by 
these methods.


\section{Superembedding Method}
 We consider a superembedding $f:M\rightarrow \unM$.
 Our index conventions are as follows; coordinate
 indices are taken from the middle of the alphabet with capitals
 for all, Latin for bosonic and Greek for fermionic, $M=(m,\m)$,
 tangent space indices are taken in a similar fashion
 from the beginning of the
 alphabet so that $A=(a,\a)$. The distinguished tangent space
 bases are related to coordinate bases by means of the
 supervielbein, $E_M{}^A$, and its inverse $E_A{}^M$. Coordinates
 are denoted $z^M=(x^m,\th^{\m})$. We use exactly the same notation
 for the target space but with all of the indices underlined.
 Target space forms are written with an underline, e.g. $\unH$. Their 
 pullbacks are written without an underline, $f^{*} \unH = H$.

 The embedding matrix is the derivative of $f$ referred to the
 preferred tangent frames, thus
\be
E_{A}{}^{\unA} = E_{A}{}^{M} \del_{M} z^{\unM} E_{\unM}{}^{\unA} \label{embedding matrix}
\ee
This tells us how to pull back target space forms onto the worldvolume,
\be
f^* E^{\unA} = E^A E_{A}{}^{\unA}.
\ee
The basic embedding condition is
\be
E_{\a}{}^{\una} = 0.
\ee
This condition in general gives constraints on the superfields describing the worldvolume theory.
For codimension zero however it can be enforced without loss of generality as discussed in 
\cite{codzero}.

The worldvolume multiplet is described by the transverse target space coordinates 
considered as superfields on the worldvolume. For codimension zero we embed an $N=1$
superspace into an $N=2$ superspace of the same bosonic dimension. Thus, in the absence of further
constraints, an unconstrained spinor superfield describes our worldvolume multiplet.

\subsection{Space Filling Branes}

We shall give a brief review of how the superembedding approach is applied to the case of the space filling branes in
flat three and four dimensional spacetime \cite{codzero}. Our bosonic indices are the same for worldvolume and target space since we 
are considering space filling branes. Our fermionic target space indices are written 
$\ua = \a i$ where $i = 1,2$ since we embed an $N=1$ superspace into an $N=2$ superspace. In 3 dimensions $\a$ is a real,
two-component Majorana spinor index. In 4 dimensions $\a$ is a complex, two-component Weyl spinor index.
The internal index $i$ is an $SO(2)$ index for the 3 dimensional case and a $U(2)$ index for the 4 dimensional case.

\subsubsection*{Supergeometry}
Firstly we specify the form of the worldvolume derivatives in terms of target space derivatives.
As discussed in \cite{codzero} we can parametrize the odd-odd part of the embedding matrix $E_{\a}{}^{\ua}$ as follows
\be
E_\a = E_{\a 1} + h_{\a}{}^{\b}E_{\b 2} \label{embedding} 
\ee
i.e.
\be
E_{\a}{}^{\b 1} = \d_{\a}^{\b} 
\quad
\text{and}
\quad
E_{\a}{}^{\b 2} = h_{\a}{}^{\b}
\ee

The worldvolume torsion can now be calculated by pulling back the 
standard flat target space torsion. As discussed in \cite{codzero}, for the codimension zero cases it is not 
necessary to introduce a worldvolume connection. Thus the torsion tells us the algebra of derivatives on the 
worldvolume. Writing our worldvolume tangent vectors $E_A$ as $\cD_A = E_{A}{}^{M} \del_M$ we have
\be
[\cD_A , \cD_B] = - T_{AB}{}^{C} \cD_C.
\ee
This algebra is the same as that introduced in 
\cite{bg,Iv}
as the algebra one ends up with by imposing that the second supersymmetry in the $N=2$ Poincar\'e superalgebra
is realised non-linearly. A similar algebra is obtained in the case of the D-9 brane \cite{abkz}.

To describe the worldvolume multiplet, one introduces a worldvolume 2-form $\cF$ (the modified field strength).
This is constrained to satisfy the Bianchi identity, 
\be
d\cF = -H \label{wvcfbianchi} 
\ee
where $H$ is the pullback onto the worldvolume of the constant, closed target space Neveu-Schwarz 3-form, $\unH$.
To get the required worldvolume $N=1$ Maxwell multiplet one imposes the standard $\cF$-constraint 
$\cF_{\a \b} = \cF_{\a b} = 0$. The constraint $\cF_{\a \b} = 0$ tells us that we have an $N=1$ Maxwell
multiplet on the brane as well as the Goldstone fermion of the embedding. Then the constraint $\cF_{\a b} = 0$
eliminates one of these spinor superfields in terms of the other. This leaves us with just the degrees of freedom
associated with the Goldstone field.
The Bianchi identity then gives a formula for $\cF_{ab}$ in terms of the degrees of freedom of the embedding.

\subsubsection*{Green-Schwarz Action}
To obtain the Green-Schwarz action for the brane we start with the Wess Zumino term in the D-brane Lagrangian \cite{hrs}. 
We construct a $D+1$ form $W_{D+1} = G_{D+1} + G_{D-1} \cF$, where the $G$ forms are pullbacks of
constant super-invariant target space RR field strengths. This form can be written explicitly as 
$W_{D+1} = dZ_D$ where $Z_D = C_{D} + C_{D-2} \cF$ and the $C$ forms are the pullbacks of the
non-invariant target space RR potentials.

Since $W_{D+1}$ is a form of degree one higher than the body of the worldvolume the fact that it is closed 
implies it is also exact and so we can also write $W_{D+1} = dK_D$ for some $K_D$. The Lagrangian form is
\be 
L_D = K_D - Z_D \label{Lagrangianform} 
\ee
and is closed by construction.
Finally the Green-Schwarz action is defined by
\be
S_{GS} = \int d^D x \e^{m_1 ... m_D} L_{m_1 ... m_D} (x,\th = 0)
\ee
where the integration is taken over the bosonic worldvolume $M_o$.

\subsubsection*{Superfield Lagrangian and Static Gauge}
To construct the superfield action \cite{codzero} one has to make a choice of gauge.
We choose the static gauge, defined by identifying the coordinates of the brane with some of the coordinates of 
the target space thus
\begin{align}
x^{\una} &= x^a \\
\th^{\a 1} &= \th^\a \\
\th^{\a 2} &= \L^\a (x,\th).
\end{align}
From the definition of the embedding matrix we can see that our choice of coordinates implies that the field 
$h_{\a}{}^{\b}$ which we introduced in (\ref{embedding}) is the worldvolume covariant derivative of the transverse
fermion field $\L$,
\be
h_{\a}{}^{\b} = \cD_{\a} \L^{\b}.
\ee
With this choice of gauge, the embedding condition $E_{\a}{}^{\una} = 0$ implies the following for the worldvolume derivatives
\begin{align}
\cD_\a &= D_\a + \psi_{\a}{}^{a} \del_a \\
\cD_a &= B_{a}{}^{b} \del_b 
\end{align}
\begin{align}
\text{where}
\hspace{10pt}
\psi_{\a}{}^{a} & {\sim} \tfrac{i}{2} D_{\a} \L \Gamma^{b} \L B_{b}{}^{a} \\
B_{a}{}^{b} & {\sim} (\d_{b}^{a} - \tfrac{i}{2} \del_{b} \L \Gamma^{a} \L )^{-1} .
\end{align}
The precise forms of the quantities $\psi$ and $B$ depend on which dimension one considers.

Given these derivatives, the one-form bases on the brane are
\begin{align}
E^{\a} &= e^{\a} \\
E^{a}  &= (e^{b} - e^{\b} \psi_{\b}{}^{b})(B^{-1})_{b}{}^{a},
\end{align}
where $e^a$ and $e^{\a}$ denote the standard one-form bases of flat superspace
\begin{align}
e^{\a} &= d \th^{\a} \\
e^a    &= d x^a - \tfrac{i}{2} d \th^{\a} (\Gamma^{a})_{\a \b} \th^{\b}. 
\end{align}
These formulae allow one to convert form components from the basis induced by the embedding to the flat basis.

To construct the superfield Lagrangian one
considers the components of the Lagrangian form in the flat basis
$e^A$, denoted $l_{ABC...}$. Generically one finds that the component $l_{\a \b c ...}$ contains a term which can be
identified with the superfield Lagrangian. In three dimensions this is a real superfield to be integrated over the whole
worldvolume superspace and in four dimensions it is a chiral superfield to be integrated over half-superspace. 

We now go on to describe the details of the superembedding for the space filling 2 and 3 branes.

\subsection{D=3}
\subsubsection*{Supergeometry}

In three dimensions the embedding is described by 
\be
\cD_\a = D_{\a 1} + h_{\a}{}^{\b}D_{\b 2} 
\hspace{10pt}
\text{ where }
\hspace{10pt}
h_{\a \b} = k\e_{\a \b} + h_a(\c^a)_{\a \b}
\ee
for $k$ real.

The closed 3-form $\unH$ has non-zero components
\be
H_{\a i \b j c} = -i (\c_c)_{\a \b} (\t_1)_{ij},
\ee
where $\t_1$ is the first Pauli matrix.

Pulling this back to the worldvolume and solving $d\cF = -H$ with the standard $\cF$-constraint we find that $k=0$ and
\be
\cF_{ab} = \frac{2}{1 + h^2} \e_{abc} h^c. \label{3dcfexp}
\ee

\subsubsection*{Action}
The Wess-Zumino form is given by $W_4 = G_4 + G_2 \cF$ where the $G$ forms are pullbacks of the 
target space RR field strengths whose non-zero components are   
\begin{align}
G_{\a i \b j} &= -i\e_{\a \b} \e_{ij} \\
G_{\a i \b j cd} &= -i(\c_{cd})_{\a \b} (\t_3)_{ij}.
\end{align}
They satisfy the Bianchi identities 
\begin{align}
\unG_{2} &= d\unC_{1}, \\
\unG_{4} &= d\unC_{3} - \unC_{1} \unH_{3}.
\end{align}
These equations can be solved for the target space potentials $\unC_{1},\unC_{3}$ so that their 
components are only functions of $\th_2$. One can then pull them back to the worldvolume and calculate 
the components of the form $Z_3 = C_3 + C_1 \cF$. Doing this one finds in static gauge
\be
Z_{abc} = \e_{abc}Z 
\hspace{10pt}
\text{where}
\hspace{10pt}
Z = 1 + i\del_a \L \c^a \L + \tfrac{1}{2} \L^2 \del_a \L^\a \del^a \L_\a. \label{3dZ}
\ee 

To complete the Green-Schwarz action we need the kinetic term. Since $W_4$ is exact we solve $W_4 = dK_3$. 
The only non-vanishing component of $K_3$ is the totally even one $K_{abc} = \e_{abc} K$. One finds
\be
K = \frac{1-h^2}{1+h^2}
\ee
which, given the relation (\ref{3dcfexp}) between $h_a$ and $\cF_{ab}$ can be shown to be of the standard Born-Infeld form.
\be
K = \sqrt{\det (\eta_{ab} + \cF_{ab})}
\ee
The Lagrangian form $L_3$ has the top component $L_{abc} = \e_{abc}L$ where $L = K - Z$.
One converts into the coordinate basis using the worldvolume supervielbein $E_{m}{}^{a}$ to obtain the 
Green-Schwarz Lagrangian 
\be
L_{GS} = \det (E) L. \label{3dgreenschwarz} 
\ee   

The superfield Lagrangian for $D=3$ is found in the $\a \b c$ component of $L$ in the flat basis \cite{codzero} :
\be
L_o \propto (\c^c)^{\a \b} l_{\a \b c}
\hspace{10pt} 
\text{i.e.}
\hspace{10pt}
L_{GS} = D^2L_o.  
\ee

\subsection{D=4}
\subsubsection*{Supergeometry}

The embedding for $D=4$ is specified by
\be
\cD_\a = D_{\a 1} + h_{\a}{}^{\b}D_{\b 2} 
\hspace{10pt}
\text{ where }
\hspace{10pt}
h_{\a \b} = k\e_{\a \b} + h_{(\a \b )}
\ee
for $k$ complex.

The closed 3-form $\unH$ has non-zero components
\be
H_{\a i\bdt c}^{\phantom{\a i} j} = -i (\s_c)_{\a \bdt} (\t_1)_{i}{}^{j}.
\ee

We now pull $\unH$  back to the worldvolume and solve $d\cF = -H$ with the standard $\cF$-constraint.
Firstly we find that the embedding preserves chirality, i.e. that $h_{\adt}{}^{\b} = 0$, 
which in static gauge says $\bar{\cD}_{\adt} \L^{\b} = 0$. We say that the Goldstone field $\L$ is
covariantly chiral.
We also get an expression for $\cF_{ab}$ in terms of $h_{\a}{}^{\b}$.
Defining the variable $s^2 = -\tfrac{1}{2}h_{(\a \b )}h^{\a \b}$, we find that
\begin{align}
\cF_{ab} &= (\s_{ab})_{\a \b}M^{\a \b} - (\tilde{\s}_{ab})_{\adt \bdt} \bar M^{\adt \bdt} \label{4dcurlyF} \\
\text{where} 
\hspace{20pt}
M_{\a \b} &= h_{(\a \b )} X(h) \label{4dM} \\
\text{and} \hspace{20pt}
X(h) &= \frac{-(1 + k\bar k) + s^2}{(1 + k\bar k)^2 - s^2 \bar s^2}. \label{4dXfn} 
\end{align}
We also find a constraint which removes one degree of freedom from the complex scalar $k$, namely
\be
k + \bar k + k (\bar k^2 - \bar s^2) + \bar k (k^2 - s^2) = 0. \label{nonlinreality} 
\ee
This is the full form of the non-linear reality constraint on the covariantly chiral Goldstone field $\L$ if 
we make the identification $h_{\a}{}^{\b} = \cD_{\a} \L^{\b}$. This constraint was first written down
to third order in $\L$ in \cite{bg}. 
The remaining degree of freedom in $k$ is the auxiliary field of the $D=4$, $N=1$ Maxwell multiplet.

\subsubsection*{Action}
The Wess-Zumino form is given by $W_5 = G_5 + G_3 \cF$ where the $G$ forms are pullbacks of the 
target space RR field strengths whose non-zero components are   
\begin{align}
G_{\a i \bdt c}^{\phantom{\a i} j} &= -i(\s)_{\a \bdt} (\t_2)_{i}{}^{j} \\
G_{\a i \bdt cde}^{\phantom{\a i} j} &= -i\e_{cdef}(\s^f)_{\a \bdt} (\t_3)_{i}{}^{j}.
\end{align}
They satisfy the Bianchi identities 
\begin{align}
\unG_{3} &= d\unC_{3}, \\
\unG_{5} &= d\unC_{4} - \unC_{2} \unH_{3}.
\end{align}
These equations can be solved for the target space potentials $\unC_{2},\unC_{4}$ so that their 
components are only functions of $\th_2 , \bar\th_2$. One can then pull them back to the worldvolume and calculate 
the components of the form $Z_4 = C_4 + C_2 \cF$. Doing this one finds
\be
Z_{abcd} = \e_{abcd}Z 
\ee
where
\be
Z = -1 +
(iB_{a}{}^{e}\del_e \L \s^a \bar\L + \hspace{5pt} \text{c.c.} )
       -\tfrac{1}{2} (B_{b}{}^{f}B_{a}{}^{e} \del_f \L \s^{ab} \del_e \L \bar\L^2 + \hspace{5pt} \text{c.c.}).
 \label{4dZ} 
\ee 
The matrix $B_{a}{}^{b}$ is given by
\be
B_{a}{}^{b} = (\d_{a}^{b} + \tfrac{i}{2} \del_{a} \s^{b} \bar{\L} + \tfrac{i}{2} \del_{a} \bar{\L} \s^{b} \L)^{-1} .
\ee
To complete the Green-Schwarz action we need the kinetic term. Since $W_5$ is exact we solve $W_5 = dK_4$. 
The only non-vanishing component of $K_4$ is totally even one $K_{abcd} = \e_{abcd} K$. One finds
\be
K = 1 + X(h) + \bar X(h) \label{4dK}
\ee

The Lagrangian form $L_4$ has the top component $L_{abcd} = \e_{abcd}L$ where $L = K - Z$.
One converts into the coordinate basis using the worldvolume supervielbein $E_{m}{}^{a}$ to obtain the 
Green-Schwarz Lagrangian 
\be
L_{GS} = \det (E) L.
\ee   
In four dimensions one can construct a chiral Lagrangian to be integrated over half-superspace.
This chiral superfield Lagrangian is found in the $\adt \bdt cd$ component of $L$ in the flat basis \cite{codzero},
\be
L_o \propto (\tilde{\s}^{cd})^{\adt \bdt} l_{\adt \bdt cd}
\hspace{10pt}
\text{i.e.}
\hspace{10pt}
L_{GS} = D^2 L_o + \bar D^2 \bar L_o. 
\ee

\section{Equivalence with non-linear realisations}

In this section we shall show the explicit relationship of the generic superembedding approach to the PBGS method  
using linear and non-linear realisations of SUSY \cite{Iv} for the case of the space filling D-branes in 3 and 4 dimensions. 
Firstly we shall discuss the general approach to showing this equivalence and then we shall go on to give the
specific formulae for the two cases under discussion.
We shall show in particular that the
standard $\cF$-constraint is equivalent to the non-linear constraints imposed in \cite{bg,ik}.

Our method will be to construct the 2-form $\cF$ of the superembedding method in the following way.  
Firstly we introduce the $N=2$ Maxwell multiplet via an independent target space 2-form $\unF$.
We constrain $\unF$ to solve the modified target space Bianchi identity
\be
d \unF = - \unH \label{modbianchi}
\ee
with the standard constraint which gives the $N=2$ Maxwell multiplet, i.e. that the lowest component of $\unF$ is written
\be
F_{\a i \b j} \propto \e_{\a \b} \e_{ij} W \label{Maxwell} 
\ee
where $W$ is an $N=2$ scalar superfield that is real in 3 dimensions and chiral in 4 dimensions.

In both three and four dimensions the Bianchi identity (\ref{modbianchi}) imposes 
modified constraints on the $N=2$ Maxwell multiplet. These constraints are precisely the deformations of the Maxwell 
constraints imposed in \cite{bg,ik,Iv}. 

We then pull back the 2-form $\unF$ onto the worldvolume of the brane identifying $f^* \unF$ with $\cF$.
By (\ref{modbianchi}) we are guaranteed that $\cF$ defined this way satisfies the correct worldvolume Bianchi identity
(\ref{wvcfbianchi}).
The components of $\cF$ are given by 
\be
\cF_{AB} = (-1)^{A.(B+ \unB)} E_{B}{}^{\unB} E_{A}{}^{\unA} F_{\unA \unB} \vline _{\phantom{.} \th_2 = \L}.
\label{Fpullback} 
\ee
We can eliminate any target space derivatives $D_1$ in favour of $D_2$ and worldvolume derivatives $\cD$ since we have 
specified the embedding to be described by (\ref{embedding}),
\be 
\cD_\a = D_{\a 1} + h_{\a}{}^{\b} D_{\b 2}. \label{derivs} 
\ee
Having evaluated the components of $\cF$ in terms of $W$, $D_2 W$ and $D_2^2 W$, we then impose the standard
$\cF$-constraint $\cF_{\a \b} = \cF_{\a b} = 0$.

These constraints imply 
\begin{align}
W \vline _{\phantom{.} \th_2 = \L} &= 0 \label{Wiszero} \\
D_{\a 2} W \vline _{\phantom{.} \th_2 = \L} &= 0. \label{DWiszero}
\end{align}

If we now expand the superfield $W$ in $\th_2$ we get generically
\be
W = \phi + \th_2^\a W_\a + ... \hspace{5pt} .
\ee
We can see explicitly that the $\cF$-constraint implies some non-linear
constraints on the component fields $\phi$ and $W_\a$. These constraints are those postulated 
in the non-linear realisations framework in \cite{bg,ik} and later derived from an algorithmic procedure in \cite{Iv}.

We then go on to show that $\cF_{ab}$ defined this way is identical to the one defined via the 
superembedding method described earlier.

\subsection{D=3}
We now show the explicit relation between the superembedding method and the method of using 
linear and non-linear realisations of PBGS for the three dimensional case.
The components of $\unF$ given by (\ref{modbianchi}, \ref{Maxwell}) are
\begin{align}
F_{\a i \b j} &= i \e_{\a \b} \e_{ij} W \\
F_{\a i b} &= \e_{ij} (\c_b)_{\a \b} D^{\b j}W \\
F_{ab} &= -i \e_{abc}(\c^c)^{\a \b} D_{\a 1} D_{\b 2}W
\end{align} 
where $W$ is a real $N=2$ scalar superfield satisfying the deformed Maxwell constraints of \cite{ik,Iv}
\begin{align}
(D_{\a 1} D_1^{\a} - D_{\a 2} D_2^{\a}) W &= -2i \label{modconstraint} \\
D_{\a 1} D_2^{\a} W &= 0 .
\end{align}
If we expand $W$ in terms of $\th_2$ these constraints imply
\be
W = \phi +i \th_2^\a W_\a -i\th_2^2 (1 - iD_1^2 \phi).
\ee

Next pull back the target space 2-form $\unF$ onto the worldvolume defining $f^* \unF = \cF$.

The components of $\cF$ are 
\begin{align}
\cF_{\a \b} &= -2i h_{(\a \b )} W \\
\cF_{\a b} &= (\c_b)_{\a \b} D_2^{\b}W + h_{\a}{}^{\c} (\c_b)_{\c \b} \bigl( \cD^\b W 
              - h^{\b \d} D_{\d 2} \bigr) W +i \cD_b \L^{\a} W \\
\cF_{ab} &= -i\e_{abc}(\c^c)^{\a \b} \bigl( \cD_\a  D_{\b 2} W - h_{\a}{}^{\c} D_{\c 2} D_{\b 2} W \bigr) 
            - 2 \Bigl( \cD_{[a} \L^\a (\c_{b]})_{\a \b} \bigl(\cD^\d W - h^{\d \e} D_{\e 2} W \big) \Bigr) 
\end{align}
where all terms are evaluated at $\th_2 = \L$.
We have written the derivatives $D_{\a 1}$ in the above expression in terms of $D_{\a 2}$ and 
worldvolume derivatives $\cD_\a$.

We next impose the standard $\cF$-constraint on this object, i.e. we set $\cF_{\a \b} = \cF_{\a b} =0$.

In terms of $\phi$ and $W_\a$ these constraints are precisely those of the PBGS method \cite{ik,Iv}
\begin{align}
\phi + i\L^\a W_\a -i\L^2 (1-iD^2\phi) &= 0 \\
iW_\a +i\L_\a (1-iD^2\phi) +\tfrac{i}{2} \L^\b \del_{\a \b} \phi -\tfrac{1}{2} \L^2 \del_{\a \b} W^\b &= 0.
\end{align} 

These give the relations
\begin{align}
\phi &= -i\L^2 (1 - iD^2 \phi) \label{phireln} \\ 
W_\a &= -\L_\a (1 - iD^2 \phi).
\end{align}

We can solve (\ref{phireln}) to get
\be
\phi = \frac{-i\L^2}{1 + D^2 \L^2}.
\ee

We can now use our relations for $\phi$ and $W_\a$ to check the expression for $\cF_{ab}$.
Using the relations (\ref{Wiszero},\ref{DWiszero},\ref{derivs}) we can see that
\be
\cF_{ab} = i\e_{abc}(\c^c)^{\a \b} h_{\a}{}^{\c}D_{\c 2} 
D_{\b 2} W \vline _{\phantom{.} \th_2 = \L}. \label{curlyFab}
\ee

We know from the $\cF$-constraint (\ref{Wiszero},\ref{DWiszero}) that on the brane we can replace the 
modified Maxwell constraint(\ref{modconstraint}) with
\be
D_{\a 2} D_2^\a W = \frac{2i}{1 + h^2}. \label{relWtoh}
\ee
Substituting this into (\ref{curlyFab}) we do indeed find the same expression for $\cF_{ab}$ as we had 
from the superembedding approach (\ref{3dcfexp}).

This is as expected since it is easy to see that $\cF$ has to be unique. Consider $\cF$ and $\cF^{\prime}$ 
satisfying 
\be
d \cF = d \cF^{\prime} = - H
\ee 
and both satisfying the standard $\cF$-constraint $\cF_{\a \b} = \cF_{\a b} = 0$.
Then $P = \cF - \cF^{\prime}$ satisfies 
\be
dP = 0 \label{dxequals0} 
\ee
and $P_{ab}$ is the only non-zero component of $P$.
The $\a \b c$ component of (\ref{dxequals0}) then implies 
\be
T_{\a \b}{}^{c} P_{cd} = 0 
\hspace{10pt}
\text{i.e.}
\hspace{10pt}
P=0.
\ee
Thus $\cF$ defined by $d\cF = -H$ and the standard $\cF$-constraint is unique.

\subsection{D=4}

We proceed in the same manner for four dimensions. The components of $\unF$ given by (\ref{modbianchi},\ref{Maxwell}) are
\begin{align}
F_{\a i \b j} &=  \e_{\a \b} \e_{ij} \bar W \\
F_{\a i \bdt}^{\phantom{\a i} j} &= 0 \\
F_{\a i b} &= i\e_{ij} (\s_b)_{\a \adt} \bar D^{\adt j} \bar W \\
F_{ab} &= -\tfrac{1}{2}
\bigl(
(\s_{ab})^{\a \b} D_{\a i}D_{\b j} W \e^{ij} 
+ (\tilde{\s}_{ab})^{\adt \bdt} \bar D_{\adt}^{i} \bar D_{\bdt}^{j} \bar W \e_{ij}
\bigr) 
\end{align}
where $W$ is a chiral $N=2$ scalar superfield satisfying the deformed Maxwell constraints of \cite{bg,Iv}
\begin{align}
D_{\a 2} D_2^{\a} W + \bar D_{\adt}^{1} D^{\adt 1} \bar W &= -2 \label{4dmodconstraint} \\
D_{\a 1} D_2^{\a} W - \bar{D}_{\adt}^{1} \bar{D}^{\adt 2} \bar{W} &= 0 \\ 
\bar D_{\a}^{i} W &= 0.
\end{align}

If we expand $W$ in terms of $\th_2$ these constraints imply
\be
W = \phi + \th_2^\a W_\a 
   +\th_2^2 (1 - \bar D_1^2 \bar \phi) 
   -\tfrac{i}{2} \th_{2}^{\a} \bar \th_{2}^{\adt} \del_{\a \adt} \phi  
   +\tfrac{i}{2} \th_{2}^{2} \bar \th_{2}^{\adt} \del_{\a \adt} W^{\a}
   -\tfrac{1}{8} \th_{2}^{2} \bar \th_{2}^{2} \square \phi
\ee
where $\phi$ and $W_\a$ are $N=1$ superfields which are chiral in the $\th_1$ direction.

Next pull back the target space 2-form $\unF$ onto the worldvolume defining $f^* \unF = \cF$.
The components of $\cF$ are
\begin{align}
\cF_{\a \b}   &= -2 h_{(\a \b )} \bar W \\
\cF_{\a \bdt} &= 0 \\
\cF_{\a b}    &= i(\s_b)_{\a \adt} \bar D^{\adt 2} \bar W 
               - ih_{\a}{}^{\c} (\s_b)_{\c \adt} 
               \bigl( 
               \bar{\cD}^{\adt} \bar W - \bar{h}^{\adt \bdt} \bar D_{\bdt 2} W 
               \bigr) 
               + \cD_b \L^{\a} \bar W \\
\cF_{ab} &= (\s_{ab})^{\a \b} h_{\a \b} D_2^2 W - (\tilde{\s}_{ab})^{\adt \bdt} \bar h_{\adt \bdt} \bar D_2^2 \bar W 
\end{align}
where all terms are evaluated at $\th_2 = \L$.
We have written the derivatives $D_{\a 1}$ in the above expression in terms of $D_{\a 2}$ and 
worldvolume derivatives $\cD_\a$.

We next impose the standard $\cF$-constraint on this object, i.e. we set $\cF_{\a \b} = \cF_{\a b} =0$.
In terms of $\phi$ and $W_\a$ these constraints are precisely those of the PBGS method \cite{bg,Iv}
\begin{align}
0 &= \phi + \L^\a W_\a +\L^2 (1 - \bar D^2 \bar \phi)  -\tfrac{i}{2} \L^{\a} \bar \L^{\adt} \del_{\a \adt} \phi 
    +\tfrac{i}{2} \L^{2} \bar \L^{\adt} \del_{\a \adt} W^{\a}  -\tfrac{1}{8} \L^{2} \bar \L^{2} \square \phi 
\label{4dphiconstraint} \\ 
0 &= W_\a - \L_\a (1 - \bar D^2 \bar \phi) - i \bar{\L}^{\adt} \del_{\a \adt} \phi
     -\tfrac{i}{2} \L_{\a} \bar{\L}^{\adt} \del_{\b \adt} W^{\b} 
     +\tfrac{i}{2} \L^{\b} \bar{\L}^{\adt} \del_{\a \adt} W_{\b} \notag \\
  &-\tfrac{1}{4} \L_\a \bar \L^2 \square \phi +\tfrac{i}{2} \L^2 \bar \L^{\adt} \del_{\a \adt} \bar D^2 \bar \phi
     +\tfrac{1}{8} \L^2 \bar \L^2 \square W_\a .
\label{4dWconstraint}  
\end{align} 

One can show that $\cF_{ab}$ defined this way agrees with (\ref{4dcurlyF}, \ref{4dM}, \ref{4dXfn}).
We know from (\ref{Wiszero},\ref{DWiszero}) that on the brane we can replace the constraint
(\ref{4dmodconstraint}) with
\be
D_2^2 W \vline _{\phantom{.} \th_2 = \L} = -\frac{1 + \bar k^2 - \bar s^2}{1 - s^2 \bar s^2}.
\ee
Employing the nonlinear reality constraint (\ref{nonlinreality}) one finds 
\be
D_2^2 W \vline _{\phantom{.} \th_2 = \L} = X(h) \label{DsquaredWisX},
\ee
which gives agreement with the expression for
$\cF_{ab}$ from the superembedding approach (\ref{4dcurlyF}, \ref{4dM}, \ref{4dXfn}).

Again this is as expected because we can use a similar argument to that given in the three dimensional
case to show that $\cF$ defined by $d\cF= -H$ and the $\cF$-constraint is unique.

After a little algebra the constraints (\ref{4dphiconstraint}, \ref{4dWconstraint}) imply
\be
\phi = \frac{W^2}{1- \bar{D} \bar{\phi}}.
\ee

This is the relation first postulated in \cite{bg} and later derived algorithmically in \cite{Iv}. It was shown \cite{bg} that it 
can be used together with its complex conjugate to show that $\phi$ agrees with a particular form of the 
$N=1 , D=4$ supersymmetric Born-Infeld superfield Lagrangian first constructed in \cite{cf}.

\section{Equivalence of the actions}
In this section we give a proof that the actions defined via the superembedding method are equivalent to those
constructed within the non-linear realisations framework. We construct the Lagrangian form of the superembedding approach
in the target space so that it agrees with that constructed in section 2 upon pullback to the worldvolume (i.e. the 
space defined by $\th_2 = \L(x,\th_1)$. We then note that the Lagrangian of the non-linear realisations framework can be 
obtained from the pullback of the same target space Lagrangian form to the space defined by $\th_2 = 0$. The two are thus 
related by an odd diffeomorphism of the target space. Due to the fact that the Lagrangian form is closed we can see that
the actions defined by the integrals of the pullbacks are invariant under target space diffeomorphisms and hence the two 
actions are equal.
 
\subsection{PBGS Action}
In \cite{bg,ik} it was observed that the leading component of the $N=2$ Maxwell field $W$ has the correct variation under the 
non-linear supersymmetry to be a candidate superfield Lagrangian.
For the three-dimensional case we can see this in the following way. Recalling that the $\th_2$ expansion of
$W$ is
\be
W = \phi + \th_{2}^{\a} W_{\a} + ...
\ee
we can see that the variation of $\phi$ under a $\th_2$ translation with parameter $\eta$ is
\be
\d_{\eta} \phi = \eta^{\a} W_{\a} .
\ee
This implies that $D^2 \phi$ varies only by a total derivative
\be
\d_{\eta} D^2 \phi = \eta^{\a} D^2 W_{\a} = \eta^{\a} \del_{\a \b} W^{\b}
\ee
and hence the action 
\be
S = \int d^3 x D^2 \phi \label{PBGSAction}
\hspace{10pt}
\propto
\hspace{10pt}
\int d^3 x d^2 \th \phi
\ee
is invariant.

Similar relations show the invariance of the the four dimensional action
\be
S = \int d^4 x (D^2 \phi + \bar{D}^2 \bar{\phi})  
\hspace{10pt}
\propto 
\hspace{10pt}
\int d^4 x d^2 \th \phi + \text{(c.c)}.
\ee
Thus in both cases $\phi$, the leading component of $W$, certainly has the correct variation to be a candidate superfield Lagrangian 
for the space filling brane.

\subsection{Target space approach}
To show that these actions agree with those defined via the superembedding approach we first construct a target space Lagrangian form.
Since we consider space filling branes we employ precisely the same argument that was used to construct the worldvolume Lagrangian forms.
We can promote the forms $W_{D+1}$ and $Z_D$ to the target space using the fact that we have defined a modified field strength $\unF$ on 
the target space:
\begin{align}
\underline{W}_{D+1} &= \unG_{D+1} + \unG_{D-1} \unF \\
\underline{Z}_D     &= \unC_{D}   + \unC_{D-2} \unF.
\end{align}
The fact that our branes are space filling means that $\underline{W}_{D+1}$ is a form of degree one higher than the body of the target
space. The fact that it is closed therefore implies it is exact and we can find a $D$-form $\underline{K}_D$ such that 
$d\underline{K}_D = \underline{W}_{D+1}$. The target space Lagrangian form $\underline{L}_{D} = \underline{K}_D - \underline{Z}_D$ is then
closed by construction.

When performing the superembedding, we consider the map 
\be
f : \cM \longrightarrow \underline{\cM}, 
\hspace{10pt}
f : (x,\th ) \longmapsto \bigl(x, \th , \L(x,\th ) \bigr). \label{braneembedding}
\ee

Now we will also want to consider the map
\be
i : \cM \longrightarrow \underline{\cM},
\hspace{10pt}
i : (x,\th ) \longmapsto \bigl(x, \th , 0 \bigr). \label{zeroembedding}
\ee

The pullback of $\underline{L}_D$ to the brane (i.e. Im $f$) will coincide with the worldvolume Lagrangian form we defined previously
(\ref{Lagrangianform}) in our discussion of the superembedding approach.
Also, the top component of the pullback of $\underline{L}_D$ to the space defined by $\th_2 = 0$ (i.e. Im $i$) will coincide with the 
Lagrangian defined via the PBGS method. We will now show this in both cases.   

\subsubsection*{D=3}
The non-zero components of the form $\underline{K}_3$ which satisfies $d \underline{K}_3 = \underline{W}_4$ are
\begin{align}
K_{\una \unb \unc} &= -i \e_{abc} (D_{1}^{2}W + D_{2}^{2}W) \\
K_{\a i \unb \unc} &= \e_{bcd} (\c^d)_{\a \b} D_{i}^{\b} W  \\
K_{\a i \b j \unc} &= -i (\c_c)_{\a \b} \d_{ij} W           
\end{align} 
We can write the top component as $K_{\una \unb \unc} = \e_{abc} \underline{K}$.
Using the modified Maxwell relation on W (\ref{modconstraint}) we find
\be
\underline{K} = -2i D_{2}^{2}W - 1 = -2i D_{1}^{2}W + 1 .
\ee

Upon pullback to the brane, $\th_2 = \L$, one can see that only the top component contributes due to the $\cF$-constraint which
says $W \vline _{\phantom{.} \th_2 = \L} = 0$ and $D_{\a 2} W  \vline _{\phantom{.} \th_2 = \L} = 0$. The pullback 
$f^* \underline{K}_3 = K_3$ therefore has only the $abc$ component.
Employing the relation (\ref{relWtoh}) we can see that the only non-zero component of the pullback $K_3$ is then
\be
K_{abc} = \e_{abc} \frac{1 - h^2}{1 + h^2}
\ee 
which agrees with the expression for $K_3$ derived from the worldvolume approach.

Alternatively we can pull back the form $\underline{L}_3 = \underline{K}_3 - \underline{Z}_3$ to the space defined by 
$\th_2 =0$. We denote this pullback with $i^* \underline{L}_3 = L_3^0$. The top component of $L_3^0$ is
\be
L^0_{abc} = \e_{abc} (-2i D^2 \phi )
\ee
which agrees with the expression for the PBGS Lagrangian (\ref{PBGSAction}).

We can therefore view the target space Lagrangian form as the parent for both actions.

\subsubsection*{D=4}
The non-zero components of the 4-form $\unK_{4}$ which satisfies $d \unK_{4} = \unW_{5}$ are given by
\begin{align}
K_{\una \unb \unc \und} &= \tfrac{1}{2} \e_{abcd} (D_1^2 W + D_2^2 W + \text{c.c}) \\
K_{\a i \unb \unc \und} &= - i \e_{abcd} (\s^a)_{\a \adt} \bar{D}^{\adt i} \bar{W}    \\
K_{\a i \b j \unc \und} &= - 2i \d_{ij} (\s_{cd})_{\a \b} \bar{W} 
\end{align}
We can write the top component as $K_{\una \unb \unc \und} = \e_{abcd} \unK$.
Using the modified Maxwell relation on W for four dimensions (\ref{4dmodconstraint}) we find
\be
\unK = D_1^2 W + \bar{D}_1^2 \bar{W} - 1 = D_2^2 W + \bar{D}_2^2 \bar{W} + 1 .
\ee

Pulling back to the brane only the top component contributes due to the $\cF$-constraint. The only non-zero component of the
pullback $f^* \unK_4 = K_4$ is the purely bosonic one,
\be
K_{abcd} = \e_{abcd} (1 + D_2^2 W + \bar{D}_2^2 \bar{W}) = 1 + X(h) +\bar{X}(h),
\ee
again giving agreement with the worldvolume approach.

If we pull $\unL_4 = \unK_4 - \unZ_4$ back to $\th_2 = 0$, defining $L_4^0 = i^* \unL_4$, we find the top component of $L_4^0$ to be
\be
L^0_{abcd} = \e_{abcd} (D^2 \phi + \bar{D}^2 \bar{\phi}),
\ee
giving agreement with the PBGS Lagrangian.

\subsection{Proof of equivalence}
In both three and four dimensional cases we see that the target space Lagrangian form $\underline{L}$ can be used to describe 
both the superembedding action and the PBGS action. The PBGS action can be written as
\be
\underset{\cM_o}{\int} e^* i^* \underline{L} ,
\ee
where $e^*$ denotes pullback to the body of $\cM$.
The superembedding action can be written as
\be
\underset{\cM_o}{\int} e^* f^* \underline{L} .
\ee
Thus the two actions are integrals of the pullbacks of the closed target space Lagrangian form to different sections of the target space.
By section we mean a bosonic submanifold that is diffeomorphic to the body of the supermanifold. 

The two integrals are the same by the following argument.

\subsubsection*{Claim} 
Consider a supermanifold $\cM$ with a $D$-dimensional body $\cM_o$. Consider the two sections $e$ , $s$ of $\cM$ defined in each
coordinate patch by
\begin{align}
e &: \cM_o \longmapsto \cM
\hspace{10pt}
;
\hspace{10pt}
e : (x^a) \longmapsto (x^a,0). \\ 
s &: \cM_o \longrightarrow \cM 
\hspace{10pt}
;
\hspace{10pt}
s : (x^a) \longmapsto (x^a,\xi^{\a}(x)).
\end{align}

Suppose further that $L$ is a closed $D$-form on $\cM$. Then 
\be
\underset{\cM_o}{\int} s^* L = \underset{\cM_o}{\int} e^* L.
\ee
This is closely related to the idea of rheonomy in the group manifold approach\cite{dftn}.
 
\subsubsection*{Proof}
Define the one parameter family of sections $s_t$ ($t$ real) in each patch by
\be
s_t : \cM_o \longrightarrow \cM
\hspace{10pt}
;
\hspace{10pt}
s_t : (x^a) \longmapsto (x^a, t \xi^{\a}).
\ee
In particular $s_0 = e$ and $s_1 = s$.

We also need the one parameter family of diffeomorphisms of $\cM$, generated by the odd vector field $\xi = \xi^{\a} D_{\a}$. 
These are denoted by $\s_t$ and defined in a patch by
\be
\s_t : \cM \longrightarrow \cM
\hspace{10pt}
;
\hspace{10pt}
\s_t : (x^a, \th^{\a} ) \longmapsto (x^a, \th^{\a} + t \xi^{\a}).
\ee

We have the composition rule
\be
s_{t+r} = \s_r \cdot s_t.
\ee

The claim will be true if 
\be
S(t) = \underset{\cM_o}{\int} s_t^* L
\ee
is independent of $t$. Differentiating with respect to $t$, we have
\begin{align}
\frac{dS}{dt} &= \lim \frac{1}{\e} 
\Bigl( \intMo s_{t+\e}^* L - \intMo s_t^* L \Bigr) \\
&= \lim \frac{1}{\e} \intMo \bigl( s_{t+\e}^* L -  s_t^* L \bigr) 
\end{align}
Using the composition rule and the fact that $(f \cdot g)^* = g^* \cdot f^*$, we have
\begin{align}
\frac{dS}{dt} &= \lim \frac{1}{\e} \intMo \bigl( (\s_{\e} \cdot s_t)^* L - s_t^* L \bigr) \\
              &= \lim \frac{1}{\e} \intMo \bigl( s_t^* \cdot \s_{\e}^* L - s_t^* L \bigr) 
\end{align}
Pulling out the pullback of the section $s$ and taking the limit inside the integral, we then have
\begin{align} 
\frac{dS}{dt} &= \lim \frac{1}{\e} \intMo s_t^* \bigl(\s_{\e}^* L - L \bigr) \\
              &= \intMo s_t^* \underset{\xi}{\cL} L .
\end{align}
The Lie derivative $\underset{\xi}{\cL} L$ can be written
\be
\underset{\xi}{\cL} L = d i_{\xi} L + i_{\xi} d L.
\ee
Since $L$ is closed this a total derivative and thus the above integral vanishes.
Hence 
\be
\frac{dS}{dt} = 0.
\ee
Thus we know that $S(t)$ is constant and hence $S(0) =S(1)$ which proves the claim. Thus the integral of the pullback of a closed
$D$-form is independent of the choice of section.
 
This claim applied to the case under consideration shows that the superembedding action and PBGS action are the same.

\section{Conclusion}

In this paper we have discussed the relationship between the superembedding approach to constructing 
brane worldvolume Lagrangians and the method of using linear and non-linear realisations of partially broken 
supersymmetry. We have focused in detail on the space filling branes in 3 and 4 dimensions.
In particular the starting assumptions of the PBGS method all have a geometrical interpretation
in terms of superembeddings. The algebra of derivatives is that induced by the pullback onto the worldvolume
of the standard flat target space torsion. We have shown that the introduction of an $N=2$ Maxwell multiplet 
satisfying modified
constraints can be understood in terms of the D-brane like worldvolume Bianchi identity $d \cF = - H$. One 
introduces an independent target space 2-form $\unF$ which satisfies the corresponding target space identity
$d \unF = - \unH$. With the standard Maxwell assumptions about the odd-odd component of this form, this gives
the $N=2$ Maxwell multiplet, with modified constraints in the target space. The covariant non-linear constraints
one has to impose in the PBGS approach to get the correct multiplet are none other than the manifestly covariant 
$\cF$-constraint.
With these identifications we have shown that the superfield Lagrangians defined by the superembedding method
are precisely equivalent to those invariants constructed via the PBGS method.

It would be interesting to see if the geometrical ideas of representing PBGS outlined here could be used
to relate the supersymmetric, non-abelian Born-Infeld action constructed in \cite{s.k} to a superembedding.
This might give some insight into what the correct brane-like form of the non-abelian Born-Infeld action is.

We emphasise that the $N=2$ Maxwell multiplet in the target space contains no degrees of freedom after imposing the 
$\cF$-constraint; they are all related to the Goldstone field of the embedding (i.e. the transverse target space
fermionic coordinate). Thus it is not clear that one has to have an $N=2$ Maxwell multiplet in the target space 
at all. One might be able to take any multiplet described by a 2-form satisfying $d \unF = - \unH$. On pulling this 
back and imposing the standard $\cF$-constraint on $\cF = f^{*} \unF$ one is guaranteed to have the $N=1$ Maxwell 
multiplet on the worldvolume. This may suggest a generalization to, say, the space filling brane in 6 dimensions where
an $N = (2,0)$ Maxwell multiplet does not exist. 
One might also be tempted to look at branes in curved backgrounds. Recently an action for a membrane in AdS4 was constructed from the 
non-linear realisations perspective \cite{adsmem}.

\section*{Appendix : Target space Ramond Potentials}

\subsection*{D=3}
The $D=3$ Ramond potentials $\unC_1$ ,$\unC_3$ solve the Bianchi identities,
\begin{align}
\unG_{2} &= d\unC_{1}, \\
\unG_{4} &= d\unC_{3} - \unC_{1} \unH_{3}.
\end{align}
Their non-zero components are : for $\unC_1$,
\begin{align}
C_{abc} &= \e{abc}  \\
C_{\a 2 bc} &= i(\c_{bc}) \th^{\b 2} \\
C_{\a 1 \b 1 c} &= (\c_c)_{\a \b} (\th^2)^2 \\
C_{\a 2 \b 2 c} &= - (\c_c)_{\a \b} (\th^2)^2 , 
\end{align}
and for $\unC_1$,
\be
C_{\a 1} = i \th_{\a}^{2}.
\ee

\subsection*{D=4}
The $D=4$ Ramond potentials solve
\begin{align}
\unG_{3} &= d\unC_{3}, \\
\unG_{5} &= d\unC_{4} - \unC_{2} \unH_{3}.
\end{align} 
Their non-zero components are : for $\unC_4$,
\begin{align}
C_{abcd} &= - \e_{abcd} \\
C_{\a 2 bcd} &= -i \e_{bcde} (\s^e)_{\a \bdt} \bar{\th}_{2}^{\bdt} \\
C_{\a 1 \b 1 cd} &= 2i (\s_{cd})_{\a \b} (\bar{\th}_2)^2 \\
C_{\a 2 \b 2 cd} &= -2i (\s_{cd})_{\a \b} (\bar{\th}_2)^2,
\end{align}
and for $\unC_2$,
\begin{align}
C_{\a 1 b} &= (\s_b)_{\a \bdt} \bar{\th}_{2}^{\bdt} \\
C_{\a 1 \b 2} &= i \e_{\a \b} (\bar{\th}_2)^2,
\end{align}
together with those obtained by complex conjugation.



\begin{thebibliography}{99}

\bibitem{d.s}
D. Sorokin, 
{\sl Superbranes and Superembeddings} Phys. Report {\bf 329} (2000) 1-101,
hep-th/9906142.

\bibitem{cs}
C. S. Chu and E. Sezgin,
{\sl M-Fivebrane from the Open Supermembrane} JHEP {\bf 9712 } (1997) 001,
hep-th/9710223.

\bibitem{chs}
C. S. Chu, P. S. Howe and E. Sezgin,
{\sl Strings and D-branes with Boundaries} Phys. Lett. {\bf B428 } (1998) 59-67, 
hep-th/9801202.

\bibitem{chsw}
C. S. Chu, P. S. Howe, E. Sezgin and P. C. West,
{\sl Open Superbranes} Phys. Lett. {\bf 429 } (1998) 273-280,
hep-th/9803041.

\bibitem{codzero}
J.M. Drummond and P.S. Howe,
{\sl Codimension zero superembeddings},
Class. Quantum Grav. {\bf 18} (2001) 4477-4492,
hep-th/0103191.

\bibitem{b.p.p.s.t}
I. Bandos, P. Pasti, A. Pokotilov, D. Sorokin, M. Tonin, 
{\sl The space filling Dirichlet 3-brane in N=2, D=4 Superspace},
Nucl.Phys.Proc.Suppl. 102 (2001) 18-25,
hep-th/0103152.

\bibitem{pst}
P. Pasti, D. Sorokin and M. Tonin,
{\sl Superembeddings, Partial Supersymmetry Breaking and Superbranes},\
Nucl.Phys. B591 (2000) 109-138,
hep-th/0007048; 
{\sl Geometrical aspects of superbrane dynamics},
Fortsch.Phys. 49 (2001) 649-656,
hep-th/0011020.

\bibitem{bg}
J. Bagger and A. Galperin, {\sl New Goldstone multiplet for partially broken supersymmetry} Phys. Rev. {\bf D55} (1997) 1091-1098, 
hep-th/9608177.

\bibitem{bg2}
J. Bagger and A. Galperin, 
{\sl The tensor Goldstone multiplet for partially broken supersymmetry} Phys. Lett. {\bf B412} (1997) 296-300, 
hep-th/9707061.

\bibitem{bik}
S. Bellucci, E. Ivanov and S. Krivonos,
{\sl Partial breaking of N=1, D=10 supersymmetry},
Phys. Lett. {\bf B460} (1999) 348-358, 
hep-th/9811244; 
{\sl Partial breaking of N=4 to N=2: hypermultiplet as a Goldstone superfield}, 
Fortsch. Phys. {\bf 48} (2000) 19-24,
hep-th/9809190; 
{\sl Superworldvolume dynamics of branes from non-linear realizations}, 
Phs. Lett. {\bf B482} (2000) 233, 
hep-th/0003273.

\bibitem{bik2}
S. Bellucci, E. Ivanov and S. Krivonos, 
{\sl N=2 and N=4 supersymmetric Born-Infeld theories from non-linear realizations},
Phys. Lett. {\bf B502 } (2001) 279-290, 
hep-th/0012236;
{\sl Towards the complete N=2 superfield Born-Infeld action with partially broken N=4 supersymmetry},
Phys.Rev. D64 (2001) 025014,
hep-th/0101195.

\bibitem{ik} 
E. Ivanov, S. Krivonos,
{\sl N=1 D=4 supermembrane in the coset approach},
Phys.Lett. B453 (1999) 237-244, hep-th/9901003.

\bibitem{rt}
M. Rocek and A. A. Tseytlin,
{\sl Partial breaking of global D=4 supersymmetry, constrained superfields, and 3-brane actions},
Phys.Rev. D59 (1999) 106001,
hep-th/9811232

\bibitem{dik}
F. Delduc, E. Ivanov, S Krivonos,
{\sl 1/4 partial breaking of global supersymmetry and new superparticle actions},
Nucl. Phys. {\bf B576 }(2000) 196-218, hep-th/9912222.

\bibitem{Iv}
E. Ivanov,
{\sl Superbranes and Super Born-Infeld Theories as Nonlinear Realizations}, 
hep-th/0105210.

\bibitem{IvKa}
E. Ivanov, A. Kapustnikov,
{\sl General relationship between linear and nonlinear realisations of supersymmetry},
J. Phys. A11 (1978) 2375-2384. 

\bibitem{iklz}
E. Ivanov, S. Krivonos, O. Lechtenfeld, B. Zupnik,
{\sl Partial spontaneous breaking of two-dimensional supersymmetry},
Nucl. Phys. {\bf B 600} (2001) 235,
hep-th/0012199.

\bibitem{acghn}
T. Adawi, M. Cederwall, U. Gran, M. Holm, B. E. W. Nilsson,
{\sl Superembeddings, Non-Linear Supersymmetry and 5-branes},
Int.J.Mod.Phys. A13 (1998) 4691-4716,
hep-th/9711203.

\bibitem{pcw}
P. West,
{\sl Automorphisms, Non-linear realizations and branes},
hep-th/0001216.

\bibitem{abkz}
V. Akulov, I. Bandos, W. Kummer and V. Zima, 
{\sl D=10 super-D9-brane} Nucl.Phys. {\bf B527} (1998) 61-94,
hep-th/9802032.

\bibitem{hrs}
P.S. Howe, O. Raetzel and E. Sezgin
{\sl On brane actions and superembeddings} JHEP {\bf 9808} (1998) 011, hep-th/9804051. 

\bibitem{cf}
S. Cecotti and S. Ferrara
{\sl Supersymmetric Born-Infeld Lagrangians} Phys. Lett. {\bf B187} (1987) 335.

\bibitem{dftn}
R. D'Auria, P. Fr\'e , P. K. Townsend, P. van Nieuwenhuizen,
{\sl Invariance of actions, Rheonomy and the new minimal N = 1 supergravity in the group manifold approach}.
Annals Phys.155:423,1984 ,
CERN-TH 3495.

\bibitem{s.k}
S. Ketov
{\sl N=1 and N=2 Supersymmetric non-Abelian Born-Infeld actions from superspace}, 
Phys.Lett. B491 (2000) 207-213,
hep-th/0005265.

\bibitem{adsmem}
F. Delduc, E. Ivanov, S. Krivonos,
{\sl Partial Supersymmetry Breaking and AdS4 Supermembrane},
hep-th/0111106.

\end{thebibliography}
 \end{document}